\documentstyle[12pt,epsfig]{article}
\begin{document}
\centerline{\Large\bf Time variation of gravitational coupling constant
in all}
\vspace*{0.025truein}
\centerline{\Large\bf  dimensions: Kasner-type cosmological models}
\vspace*{0.050truein}
\centerline{Forough
Nasseri\footnote{Email: nasseri@fastmail.fm}}
\centerline{\it Institute for Astrophysics,
P.O.Box 769, Neishabour, Iran}
\begin{center}
(\today)
\end{center}
\begin{abstract}
We formulate the time variation of the gravitational coupling constant
in all dimensions. We show that the time variation of the
gravitational coupling constant is related to the time variation
of the Newton's constant in three-space dimensions and also is related
to the time variation of the volume of the extra spatial dimensions.
Our results here are based on a model-independent approach.
We then study time variation of the gravitational coupling constant
in Kasner-type cosmological models. In the case of $(4+1)$-dimensional
spacetime we show that the gravitational coupling constant
decreases within the cosmic time and the rate of its time variabilty
is of the order of the inverse of the present value of the age of
the universe.\\
\vspace*{0.035truein}
\end{abstract}

\section{Introduction}
Usually in cosmological models based on higher dimensions the problem of
dimensionality of the gravitational coupling constant is not tackled on, being
tacitly assumed to be
\begin{equation}
\label{1}
\kappa=8 \pi G,
\end{equation}
where $G$ is the Newton's gravitational constant. This is of course a
relationship being derived in $(3+1)$-dimensional spacetime.
In this paper, we first review time variation of the gravitational
coupling constant in all dimensions by using a model-independent
approach and second study time variation of gravitational coupling
constant in Kasner-type cosmological models.

The plan of this paper is as follows. In Section 2, we give a brief
review of the gravitational coupling constant in all dimensions by using a
model-independent approach. In Section 3, we use the results presented
in Section 2 and obtain the time variation of the gravitational
coupling constant in Kasner-type cosmological models.
Finally, we discuss our results and conclude in Section 4.
We will use the natural units system that sets $k_B$, $c$, and $\hbar$
are equal to one, so that $\ell_P=M_P^{-1}=\sqrt{G}$.

\section{Time variation of gravitational coupling constant
in all dimensions}

Let us here review the gravitaional coupling constant in all dimensions
(see Ref. \cite{1}).
Take the metric in $(D+1)$-dimensional spacetime in the following
form 
\begin{equation}
\label{2}
ds^2= - dt^2 +a^2(t) d \Sigma_k^2,
\end{equation}
where $d \Sigma_k^2$ is the line element for a $D$-manifold of constant
curvature $k=-1, 0, +1$, corresponding to the closed, flat and hyperbolic
spacelike sections, respectively.
Using Eq.(\ref{2}), we obtain\footnote{It is worth mentioning that
in $D$-dimensional spaces the authors of Ref.\cite{3} have considered
$R_{00}= \nabla^2 \phi$. For this reason, the results presented in
Ref.\cite{3} are not in agreement with the results presented here
and Refs.\cite{1}}
\begin{equation}
\label{3}
R_{00}= (D-2) \nabla^2_{D} \phi,
\end{equation}
where $\nabla_D$ is the $\nabla$ operator in $D$-dimensional spaces. In
$(3+1)$-dimensional spacetime, the Poisson equation is given by
$\nabla^2 \phi = 4 \pi G \rho$. Applying Gauss law for a $D$-dimensional
volume, we find the Poisson equation for arbitrary fixed dimension
\begin{equation}
\label{4}
\nabla^2_D \phi = S^{[D]} G_{(D+1)} \rho,
\end{equation}
where $G_{(D+1)}$ is the $(D+1)$-dimensional Newton's constant and
$S^{[D]}$ is the surface area of a unit sphere in $D$-dimensional
spaces
\begin{equation}
\label{5}
S^{[D]} = \frac{2 \pi^{D/2}}{\Gamma \left( \frac{D}{2} \right) }.
\end{equation}
On the other hand from Eq.(\ref{2}) we get
\begin{equation}
\label{6}
R_{00} = \left( \frac{D-2}{D-1} \right) \kappa_{(D+1)} \rho.
\end{equation}
Using Eqs.(\ref{3}-\ref{6}), we are led to the gravitational coupling
constant in $(D+1)$-dimensional spacetime\footnote{For example,
in the cases $(3+1)$, $(4+1)$ and $(5+1)$-dimensional spacetime
we have $\kappa_{(3+1)}=8 \pi G_{(3+1)}$ (i.e. $\kappa = 8 \pi G$),
$\kappa_{(4+1)} = 6 \pi^2 G_{(4+1)}$ and $\kappa_{(5+1)}=
\frac{32 \pi^2}{3} G_{(5+1)}$, respectively.}
\begin{equation}
\label{7}
\kappa_{(D+1)} = (D-1) S^{[D]} G_{(D+1)}=
\frac{2 (D-1) \pi^{D/2} G_{(D+1)}}{\Gamma \left( \frac{D}{2} \right)}.
\end{equation}
Let us now obtain a relationship between the Newton's constant in
$(D+1)$-dimensional spacetime and in the $(3+1)$-dimensional spacetime.
Using the force laws in $(D+1)$- and $(3+1)$- dimensional spacetime,
which are defined by
\begin{eqnarray}
\label{8}
F_{(D+1)} (r) &=& G_{(D+1)} \frac{m_1 m_2}{r^{D-1}},\\
\label{9}
F_{(3+1)} (r) &=& G_{(3+1)} \frac{m_1 m_2}{r^2},
\end{eqnarray}
and the $(D+1)$-dimensional Gauss law one can derive (see Ref.\cite{4})
\begin{equation}
\label{10}
G_{(3+1)}= \frac{S^{[D]}}{4 \pi} \frac{G_{(D+1)}}{V^{[D-3]}},
\end{equation}
where $V^{[D-3]}$ is the volume of $(D-3)$ extra spatial dimensions.
Now, using Eqs.(\ref{7}) and (\ref{10}) we are led to
\begin{equation}
\label{11}
\kappa_{(D+1)} = 4 \pi (D-1) G_{(3+1)} V^{[D-3]}.
\end{equation}
Time derivative of Eq.(\ref{11}) leads to
\begin{equation}
\label{12}
\frac{{\dot \kappa}_{(D+1)}}{\kappa_{(D+1)}}=
\frac{{\dot G}_{(3+1)}}{G_{(3+1)}} + \frac{{\dot V}^{[D-3]}}{V^{[D-3]}}.
\end{equation}
This equation tells us the time evolution of the gravitational
coupling constant in $(D+1)$-dimensional spacetime is related to the
time evolution of the Newton's conatant
in three-space dimensions and also to the time evolution of the volume
of $(D-3)$ extra space dimensions.
Using Ref.\cite{5},
the time variation of Newton's constant today
has an upper limit
\begin{equation}
\label{13}
\bigg| \frac{{\dot G}_{(3+1)}}{G_{(3+1)}} \bigg| \bigg|_0 < 9
\times 10^{-12} {\rm yr}^{-1}.
\end{equation}
Using this upper limit and Eq.(\ref{12}) we are led to an upper limit
for the time evolution of the gravitational coupling constant.
Our approach here for the time evolution of the gravitational coupling
constant is a model-independent approach and may be used for
cosmological models in higher dimensions.

\section{Time variation of gravitational coupling constant
in Kasner-type cosmological models}
In this section, we are going to obtain the time-variation of the
gravitational coupling constant for Kasner-type cosmological models in
all dimensions.
In the case of $(3+1)$-dimensional spacetime, the general cosmological
solutions of the vacuum Einstein equations are the Kasner solutions,
with vaccum Kasner metric
\begin{equation}
\label{14}
ds^2= dt^2-t^{2p_1} dx_1^2-t^{2 p_2} dx_2^2 - t^{2 p_3} dx_3^2,
\end{equation}
where
\begin{equation}
\label{15}
p_1+p_2+p_3=p_1^2+p_2^2+p_3^2=1.
\end{equation}
This constraint requires at least one of the $p_i$ to be negative.
Thus in $(3+1)$-dimensional spacetime, the vacuum Kasner solution is
a poor description of the universe, since it predicts contradition in
at least one dimension.
In the case of $(D+1)$-dimensional spacetime, the Kasner metric is given
by
\begin{equation}
\label{16}
ds^2=dt^2 - \sum_{i=1}^{D} \left( \frac{t}{t_0} \right)^{2p_i} (dx_i)^2.
\end{equation}
The Kasner metric is a solution to the vacuum Einstein equations provided
the Kasner conditions are satisfied by
\begin{equation}
\label{17}
\sum_{i=1}^{D} p_i= \sum_{i=1}^{D} p_i^2=1.
\end{equation}
In order to satisfy the Kasner conditions at least one of the $p_i$ must
be negative. It is possible to have three spatial dimensions expanding
isotropically and $(D-3)$ extra dimensions contracting isotropically,
with the choice \cite{6}
\begin{eqnarray}
\label{18}
p_1=p_2=p_3 &\equiv& p = \frac{3+(3D^2-12D +9)^{1/2}}{3D},\\
\label{19}
p_4=...=p_{D-3} &\equiv& q =\frac{D-3-(3D^2 -12D+9)^{1/2}}{D(D-3)}.
\end{eqnarray}
Provided that $D-3>0$, $p>0$ and $q<0$ as desired. With this choice the
metric may be written as
\begin{equation}
\label{20}
ds^2= dt^2 - a^2(t) \sum_{i=1}^{3} d{x_i}^2 - b^2(t) \sum_{i=4}^{D}
d{y_i}^2,
\end{equation}
where $x_i$ are the spatial coordinates of the three expanding dimensions,
and $y_i$ are the spatial coordinates of the $D-3$ contracting dimensions.
The two scale factors evolve as $a(t)=(t/t_0)^p$, $b(t)=(t/t_0)^q$.

To study the time variation of the gravitational coupling constant
for Kasner-type cosmological models in all dimensions, we use
\begin{equation}
\label{21}
V^{[D-3]}=b^{(D-3)}=\left( \frac{t}{t_0} \right)^{q(D-3)}.
\end{equation}
Therefore, we have
\begin{equation}
\label{22}
\frac{{\dot V}^{[D-3]}}{V^{[D-3]}}=\frac{q(D-3)}{t}.
\end{equation}
Using Eqs.(\ref{12}), (\ref{19}) and (\ref{22}), we get
\begin{equation}
\label{23}
\frac{{\dot \kappa}_{(D+1)}}{\kappa_{(D+1)}}
=\frac{{\dot G}_{(3+1)}}{G_{(3+1)}} +
\frac{D-3-(3D^2 -12D+9)^{1/2}}{Dt}.
\end{equation}
Inserting $D=4$ in Eqs.(\ref{18}) and (\ref{19}),
in the case of $(4+1)$-dimensional
vacuum Kasner metrics we have $p_1=p_2=p_3=1/2$ and $q=-1/2$.
Therefore, time variation of the gravitational coupling constant
in $(4+1)$-dimensional spacetime of Kasner-type cosmological models
is given by
\begin{equation}
\label{24}
\frac{{\dot \kappa}_{(D+1)}}{\kappa_{(D+1)}}=
\frac{{\dot G}_{(3+1)}}{G_{(3+1)}} - \frac{1}{2t}.
\end{equation}
At the present time, the above equation reads
\begin{equation}
\label{25}
\frac{{\dot \kappa}_{(D+1)}}{\kappa_{(D+1)}}\bigg|_0=
\frac{{\dot G}_{(3+1)}}{G_{(3+1)}}\bigg|_0 - \frac{1}{2t_0},
\end{equation}
where $t_0$ is the age of the universe at present time being
$t_0 \sim 10^9 {\rm yr}$.
Using Eqs.(\ref{13}) and (\ref{25}), in $(4+1)$-dimensional
spacetime the time variation of the gravitational coupling constant
to be
\begin{equation}
\label{26}
\frac{{\dot \kappa}_{(D+1)}}{\kappa_{(D+1)}}\bigg|_0  \sim
\left( 10^{-12} - 10^{-9} \right)\;{\rm yr}^{-1} 
\sim  -10^{-9}\;{\rm yr}^{-1}.
\end{equation}
This means that if we take the present value of the space dimensions
to be $D_0=4$ in Kasner-type cosmological models, the time variation of
the gravitational coupling constant is due to not only the time varition
of Newton's constant in $(3+1)$-dimensional spacetime,
$ \sim 10^{-12} {\rm yr}^{-1}$,
but also the inverse of the age of the universe today, $ \sim 10^{-9}
{\rm yr}^{-1}$. From Eq.(\ref{26}) we are led to that in Kasner-type
cosmological models and  $(4+1)$-dimensional spacetime the gravitational
coupling constant decreases within the cosmic time and the rate
of its time variability is approximately $10^{-9} {\rm yr}^{-1}$.

\section{Conclusions}
In this paper, we evaluated the time variation of the gravitational coupling
constant in all dimensions by using a model-independent approach.
We then use, our result in the case of Kasner-type cosmological
models. We conclude that in the case of $(4+1)$-dimensional spacetime
and in Kasner-type cosmological model, the time evolution of the gravitational
coupling constant is due to not only the time variation of the
Newton's constant in $(3+1)$-dimensional spacetime but also
the inverse of the age of the universe. As given in Eq.(\ref{26}),
in Kasner-type cosmological models in  $(4+1)$-dimensional spacetime 
the gravitational coupling constant decreases within the cosmic time
with a rate of order $ \sim 10^{-9} {\rm yr}^{-1}$.
Our approach in this paper may be use to study time variation of
the gravitational coupling constant for cosmological model in
higher dimensions.\\
\noindent
{\bf Acknowledgments:} F.N. thanks Hurieh Husseinian and
Ali Akbar Nasseri for noble helps.

\end{document}